\documentclass{llncs}

\usepackage{amsmath}
\usepackage{url}
\usepackage{graphicx}
\usepackage{subfig}
\usepackage{caption}
\usepackage{wrapfig}
\usepackage[font=itshape]{quoting}

\newtheorem{mydef}{Definition}

\DeclareMathOperator*{\argmin}{argmin}

\begin{document}%


\title{Towards Automated Boundary Value Testing with Program Derivatives and Search}

\titlerunning{Towards Automated Boundary Value Testing...}  
%
\author{Robert Feldt and Felix Dobslaw}
\authorrunning{Feldt and Dobslaw} 
%
\tocauthor{Robert Feldt and Felix Dobslaw}
%
\institute{Dept. of Computer Science and Engineering,\\
Div. of Software Engineering,\\
Chalmers University of Technology,\\
Gothenburg, Sweden,\\
\email{robert.feldt@chalmers.se}
}

\maketitle              

\begin{abstract}
A natural and often used strategy when testing software is to use input values at boundaries, i.e. where behavior is expected to change the most, an approach often called boundary value testing or analysis (BVA).
Even though this has been a key testing idea for long it has been hard to clearly define and formalize. Consequently, it has also been hard to automate.

In this research note we propose one such formalization of BVA by, in a similar way as to how the derivative of a function is defined in mathematics, considering (software) \textit{program derivatives}. Critical to our definition is the notion of distance between inputs and outputs which we can formalize and then quantify based on ideas from Information theory. 

However, for our (black-box) approach to be practical one must search for test inputs with specific properties. Coupling it with search-based software engineering is thus required and we discuss how program derivatives can be used as and within fitness functions.

This brief note does not allow a deeper, empirical investigation but we use a simple illustrative example throughout to introduce the main ideas.
By combining program derivatives with search, we thus propose a practical as well as theoretically interesting technique for \textit{automated boundary value (analysis and) testing}.


\end{abstract}
\section{Introduction}

Software systems increasingly govern our modern society and it is essential that we have effective and efficient ways to avoid that they contain critical faults. An old and natural way for software practitioners to think when creating software tests is to try to identify borders where the behavior of the software should change (the most). Even though such boundary value analysis (BVA) and testing (BVT) based on it is a classic technique in the software testing literature~\cite{clarke1982close}, and typically a mandatory part of relevant textbooks and certification programs~\cite{spillner2014software,bath2012software}, there has been only limited progress on how to objectively formalize and define it in a general way.

Boundary value analysis is closely related to partition analysis (PA) which divides the input domain for the software under test (SUT) into sub-domains for which we expect the behavior to be uniform for all inputs within the domains~\cite{hierons2006avoiding}. If the software behaves incorrectly on a sub-domain, the intuition is that it should fail for many or all of its elements. Further, if this holds true, we only need to test one or a few inputs per sub-domain which reduces the testing efforts many-fold. Practical experience also shows that software developers frequently introduce faults at domain borders, for example the common `off-by-one' errors~\cite{glass2001frequently}. Identifying boundaries in the input domain and adding test cases to detect faults at such boundaries is thus often an effective testing strategy.

After the initial papers by White and Cohen~\cite{white1980domain} and Clarke et al.~\cite{clarke1982close} that introduced and extended the basic method for boundary value analysis and testing\footnote{also called domain testing in several papers but less so in recent years.} it was further refined by a number of authors. Jeng and Weyuker~\cite{weyuker1991analyzing} simplified and generalized the approach to also cover discrete-valued inputs and Jeng and Forgacs~\cite{jeng1999automatic} then proposed a semi-automated approach in which a dynamic search for test inputs is combined with algebraic manipulation of the boundary conditions in order to more efficiently generate test data for BVT. 

However, a downside of all of these approaches is that they target numeric real-valued inputs. Discrete-valued inputs are sometimes supported by approximation schemes. Only recently did Zhao et al.~\cite{zhao2010automatic} consider string inputs and showed how to generate test data to better find problems at borders in code with string predicates. Their basic idea is to introduce a specific string distance metric that is adapted to how strings are typically compared (lexicographically) in string predicates.
Since software not only allow for numeric or string inputs, but frequently also complex and structured inputs, it is not obvious how BVA may be generalised or how it facilitates automated testing. 

In this research note we argue that a general and sound basis for BVA can be created by considering the information distance between test inputs and their resulting outputs. The information distance is based on Kolmogorov complexity and is thus applicable to any type of object~\cite{bennett1998information} and thereby to any type of software input. Since Kolmogorov complexity is not computable it may seem as though little has been gained. But by approximating it with compression algorithms it has been shown repeatedly and in various domains that these information theoretical metrics can reach state-of-the-art results~\cite{cilibrasi2004algorithmic,vitanyi2009normalized}. Furthermore, they can do so without any specific knowledge about the objects features, their importance, or the type of similarity to be considered.

Feldt et al.~\cite{feldt2008searching} have previously applied these metrics to measure the distance between software tests and proposed their use to search for test cases. More recently, Feldt et al.~\cite{feldt2016tsdm} generalized these results to search for diverse sets of test cases and Marculescu and Feldt~\cite{marculescu2018finding} investigated distance metrics in robustness testing. In this research note, we propose a further application of these information theoretic measures in software testing: searching for pairs of test inputs for automated boundary value testing based on the normalized compression distance.

In the following we first introduce a short, illustrative example in Section \ref{example}. Then, in Section \ref{theory}, we introduce a general formalism for the location of boundaries which is inspired by the concept of the classic single-value derivative of a function in (mathematical) calculus. In Section \ref{applied}, we apply our approach to an implementation of the illustrative example specification, and locate boundaries without source-code access, i.e. in a black-box manner. This note closes with a discussion on ways forward and in particular how to use our proposed program derivative in search-based software testing for automated BVA.

\section{Illustrative Example: Constrained Sum}
\label{example}

We focus on the general situation where there is a given specification and then some software that implements it. Further, we don't have or don't want to access the source code and we seek relevant boundaries for testing. Thus, a black-box, boundary value testing scenario. We here simply call this example (a variant of) the constrained sum problem with the specification:

\begin{quoting}
\textbf{Constrained Sum}: The software should calculate the sum of two floating input values. The result should be returned with one (significant) decimal. Negative input values are invalid, as well as any inputs or outputs larger than or equal to 6.
\end{quoting}

\begin{wrapfigure}{R}{0.3\textwidth}
\centering
\vspace{-.2cm}
\includegraphics[width=0.25\textwidth]{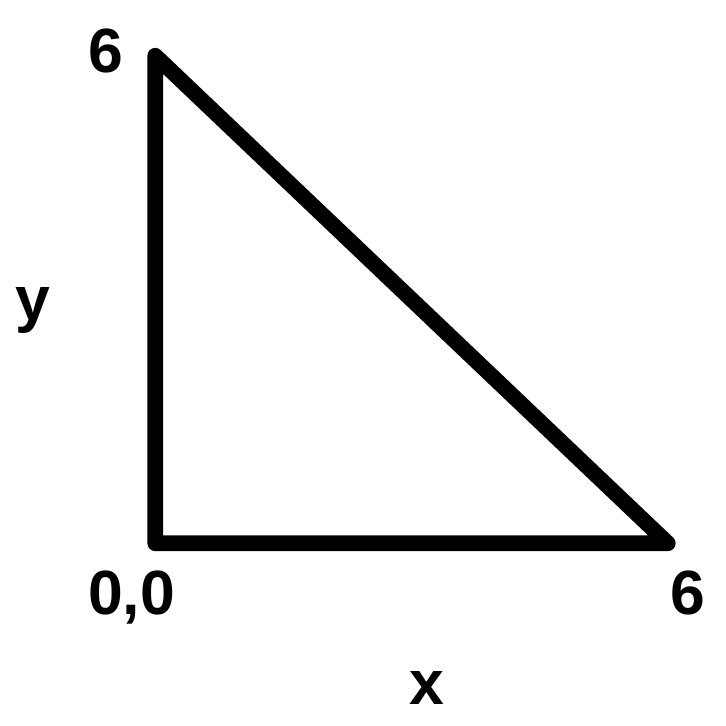}
\caption{\label{fig:specification}Boundaries as conceptualized for the constrained sum program.}
\end{wrapfigure}

Figure~\ref{fig:specification}, on page \pageref{fig:specification}, shows a conceptual picture of the boundaries the software engineer had in mind while writing and reasoning about this function for the software system.




\section{Difference Quotient and Derivative of a Program}
\label{theory}

The classic derivative from mathematical calculus is based on comparing the difference between the outputs of a function given as small a change as possible in the inputs. It is typically formally defined in terms of one point, $x$, and a delta value, $h$, which together define a second point after summation. The derivative is then the limit as the delta value goes to zero:

	$$\lim_{h\to 0} \frac{f(x+h) - f(x)}{h}$$


We argue that this, very general, idea is close to what we want to do in boundary value analysis. We need something akin to a derivative for a software program. A derivative in standard, mathematical calculus measures the sensitivity to change of a quantity (often called the function value or the dependent variable) as determined by another quantity (the independent variable). A large (absolute value of a) derivative thus indicates large sensitivity to the input, independent, variable. Detecting inputs that are highly sensitive to small changes, i.e. nearby inputs for which outputs differ a lot, would thus help us identify boundaries. It is there where we likely should spend more time testing.

However, in contrast to the continuous, single-input functions studied in mathematical calculus, programs in software can have other types than real-valued numbers as inputs and/or outputs. They also commonly have more than one input and sometimes more than one output. The latter problem can be approached in a similar way to how it is done in calculus, i.e. for functions of multiple input values we can define \textit{partial derivatives}. The former problem concerning the restricted domains is more fundamental. How can we construct a new input point from a given input point and a `delta' value, and what does the `delta' even mean, e.g. for structured input domains such as graphs, trees or databases?




In order to resolve this we take the alternative viewpoint on defining derivatives, namely the `difference quotient' over an interval~\cite{wiki:differencequotient}:

	$$DQ(a,b) = \frac{f(a) - f(b)}{a - b}$$

The derivative can now be found by letting the input $b$ go towards the input $a$. Note that the subtract operation `-' here, essentially, acts as a distance function twice, once for the outputs and once for the inputs\footnote{Subtraction also preserves directionality, however, in the following we focus purely on the distance (the absolute value) rather than on its directionality (which we argue is less clear a concept for arbitrary data types).}.
For software programs, when neither the inputs nor the outputs might be numbers, we must generalize this and allow for a general distance function for any type of data (inputs and outputs) rather than assuming we can simply use subtraction. Formally we thus define the Program Derivative as follows:

%
%

\begin{mydef}
The Program Derivative (PD) for program $P$ at input $a$, with output distance function $d_o$, and input distance function $d_i$ is

\begin{equation*}
\begin{aligned}
PD_{d_o,d_i}(a) & = PDQ_{d_o,d_i}(a, b_{min}) & = & \frac{d_o(P(a), P(b_{min}))}{d_i(a, b_{min})} &&\text{  with} \\
& b_{min} &=& \argmin_{b, b \neq a} {d_i(a, b)},&&
\end{aligned}
\end{equation*}

where $P(x)$ denotes the output of the program for input $x$.
\end{mydef}


The PD and the program difference quotient\footnote{We note that for search-based testing the PDQ, used on the right hand side of the PD definition, might be a more fruitful concept than the derivative itself since, for complex and high-dimensional data domains the closest value to another value can be ill-defined, and there can be several directions that are interesting to consider for sensitivity and rates of change (not only the one of the closest `neighbour').}, PDQ, are parameterized on two distance functions: one for the inputs and one for the outputs. They may be the same but need not be; it depends on the types of the inputs and outputs, respectively, and which of the often many possible distance functions for one type are chosen.
The most general choice of distance function is to use the Normalized Information Distance ($NID$) for both inputs and outputs since it is both universal and general and should capture any important differences~\cite{bennett1998information,vitanyi2009normalized}. We will call thisfar theoretical measure the information difference quotient (IDQ).
%
By using the `compression trick' of Cilibrasi and Vitanyi we can approximate the IDQ by substituting a compression function, $C$, for Kolmogorov complexity~\cite{cilibrasi2004algorithmic}. We thus define the Compression Difference Quotient (CDQ) of a program $P$ for inputs $a$ and $b$:

	$$ CDQ_{C}(a, b) = PDQ_{NCD_{C},NCD_{C}}(a, b) = \frac{NCD_{C}(P(a), P(b))}{NCD_{C}(a, b)} $$

where NCD is the Normalized Compression Distance~\cite{cilibrasi2004algorithmic}. However, we note that if either the inputs or the outputs are numbers, numerical vectors or matrices it may be sensible to use data-type specific distance functions. In general we thus talk about the $PDQ_{d_1,d_2}$ where $d_1$ and $d_2$ is by default NCD, but can be any chosen as any suitable distance function. 

The program derivative and its quotient thus imply whole families of concrete measures that can be instantiated and then utilized for different testing and analysis purposes. By selecting specific distance functions and calculating the quantities defined by the formulas above, we should be able to detect areas of special interest for software comprehension and quality assurance tasks.

The connection to search-based software testing seems rather direct. For complex and structured data types it might be very hard to define how to maximize and minimize the involved quantities or take `delta' steps between values. Thus, even though more exact search and optimization approaches might be useful for some programs and distance functions we can always fall back on general, black-box, meta-heuristic optimization. A good base choice might be an evolutionary search algorithm connected to a data generation framework, e.g. \cite{feldt2013finding}, but also alternative search methods can be called for~\cite{feldt2015broadening,loscher2017targeted}.


\section{Boundary Value Analysis of Constrained Sum}
\label{applied}

\begin{figure}[htbp]
\centering
\subfloat[\label{fig:triangle_boundaries_no_error}Program one.]{\includegraphics[width=.5\textwidth]{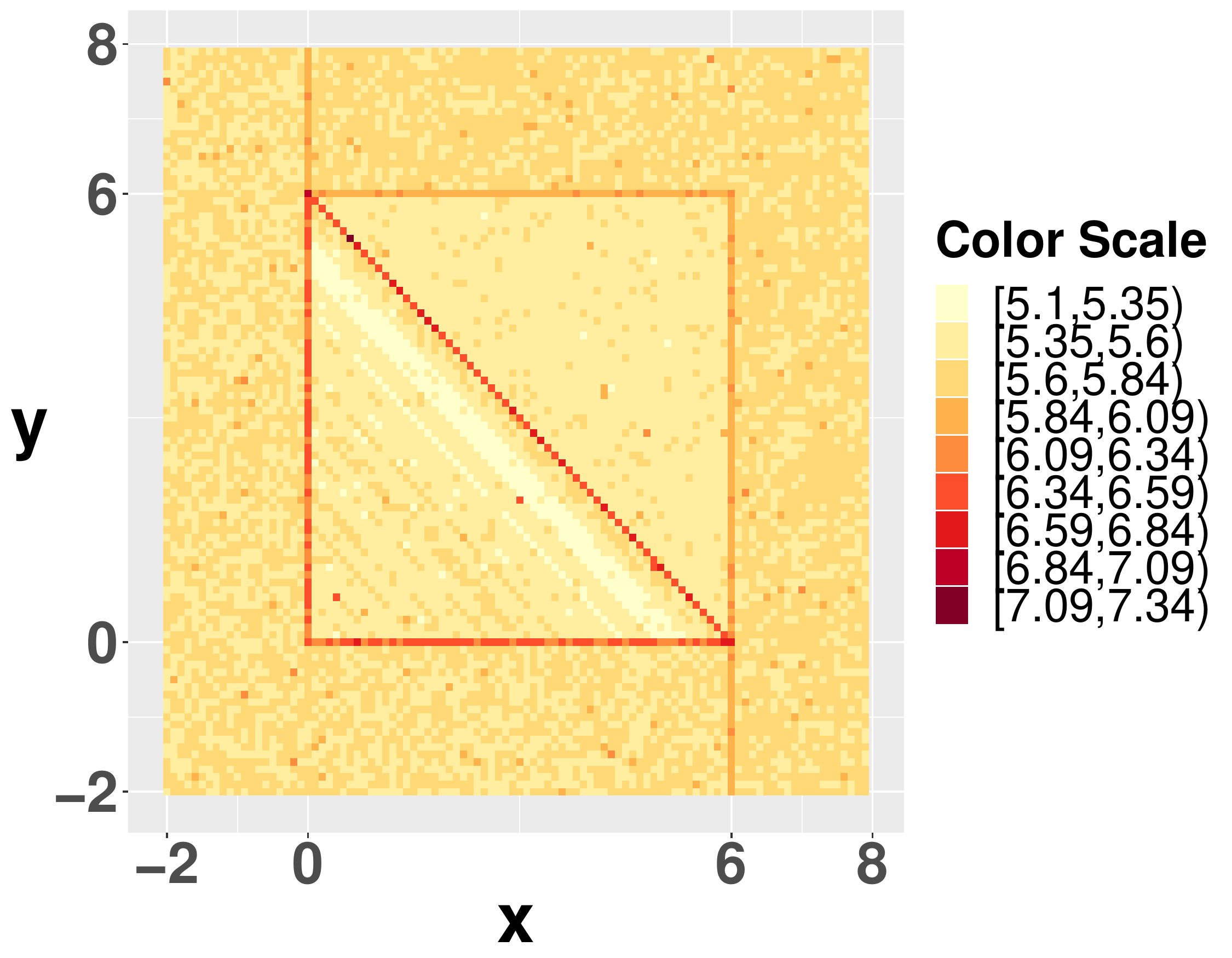}}
\subfloat[\label{fig:triangle_boundaries_error}Program two.]
 {\includegraphics[width=.5\textwidth]{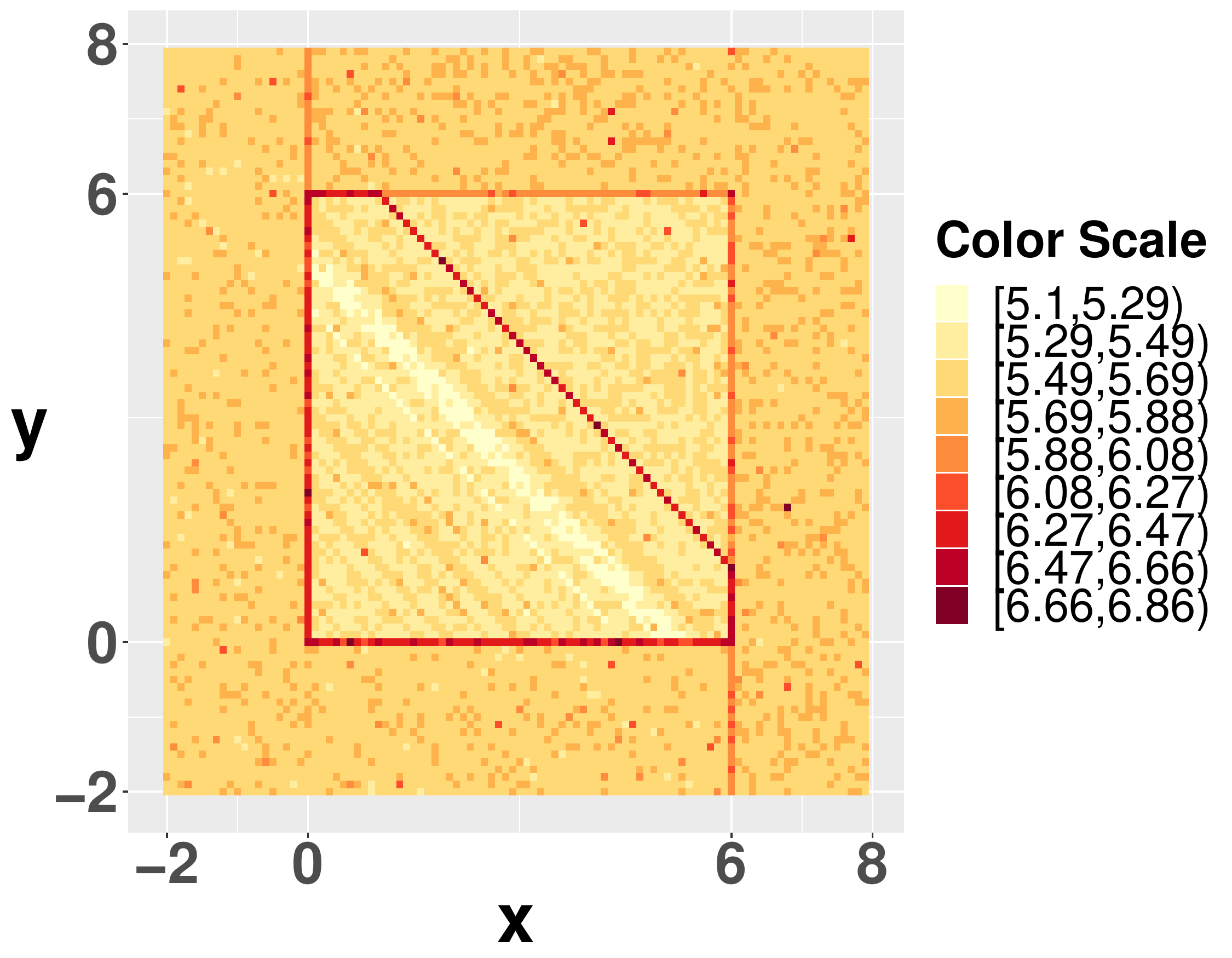}}
\caption{Heatmaps of the program derivative values for each cell of a part of the two-dimensional input space for two different implementations of the constrained sum specification from Section~\ref{example}. One program implements the specification (left) and one does not (right). Boundaries of changing behavior are clearly present.} \label{fig:triangle_boundaries}
\end{figure}

For the purpose of understanding local output differences for the software in our illustrative example from Section~\ref{example}, we applied the $CDQ_{C}$  on the grid of values in the range covered by $x,y \in [\text{-}2,8]$. For each point in a cell on a grid we then sampled a set of surrounding points, calculated $CDQ_{C}$, and selected the one with the maximal value\footnote{In an automated BVA tool we would instead have used search here.}. We then color-coded the $CDQ_{C}$ values in order to visualize the local differences in a two-dimensional plane presented in Figure \ref{fig:triangle_boundaries_no_error}. The darker the color of a pixel, the more diverse the outputs of the neighboring inputs it represents according to the applied generic measure. This way, and even without specification, we can learn local functional properties for the software system regardless of its input and output data types (since any data can be dumped to a string and a compressor applied to it). 



Figure \ref{fig:triangle_boundaries_error} shows the result of the exact same experiment for another program with the same interface. The plot is clearly dissimilar to Figure \ref{fig:triangle_boundaries_no_error} which suggests that it does not implement the specification. When looking more closely we find that the region of similar values in the center is larger, and in fact the second program allows for the sum to be larger or equal to 7. Comparing this to the conceptual image of what we expect the software to do, Figure~\ref{fig:triangle_boundaries_no_error}, it seems clear that there might be some problem with the implementation in Figure~\ref{fig:triangle_boundaries_error}. Further, we might want to reason about the significance of the `boundaries' outside of the triangle in Figure~\ref{fig:triangle_boundaries_no_error}.

\section{Discussion}
\label{discussion}

We have proposed program derivatives to detect boundaries for boundary value analysis and testing. They can be seen as a generalisation of derivatives of single-valued mathematical functions in calculus. While the mathematical derivative and differential calculus of Leibniz and Newton focus on numbers, we propose the use of information distance and the complexity trick to generalize this concept to any input and output data types. It can thus be applied to any program. While there have been some proposals for derivatives of specific types of programs such as parsers~\cite{might2011parsing} we know of no general attempts as the one proposed in this paper. 

Since this is only a brief research note there are many avenues for future work. The illustrative example we used here was chosen for simplicity and visualisation and as such takes numbers as inputs, which goes against our main motivation. However, the outputs can be of many different types (exceptions were thrown for invalid values, for example) so already on this simple example the $NCD$ allowed comparing values of different types. But future work should also explore the many different ways in which these proposed derivatives and differential quotients can be used in fitness functions and coupled with search-based testing. For example, Marculescu and Feldt~\cite{marculescu2018finding} proposed a search-based algorithm to find a border between the valid and invalid values of a program under test. We should combine this type of search to `squeeze' a border with the measures proposed in here to even find other types of borders. 

Furthermore, it is not clear that minimizing the denominator in difference quotients is the single possible goal. For constructing sets of interesting test cases we will most likely need a multi-objective formulation that combines diversity of sets of values~\cite{feldt2016tsdm} and derivatives\slash quotients. Practical work on how to select interesting and relevant distance functions for particular purposes and how to speed up distance calculations are also important and recent advances show promise~\cite{cruciani2019scalable}. 

A more conceptually intriguing area for future work would be to consider derivatives and quotients of other types of program- and test-related information. As was noted already by Feldt et al.~\cite{feldt2008searching}, all types of test-related information can be used in information distances and their approximations might have value. They also calculated distances and diversities both on inputs, state information captured in execution traces, and outputs. Alshahwan and Harman later saw promising results when using output diversity~\cite{alshahwan2012augmenting}. Nevertheless, we propose to investigate the benefits of relating different diversities and distances to each other in more ways than outlined here. For example, we can consider different partial derivatives or relating other quantities, e.g. the derivative of a program's state (output) with respect to one of its inputs (state variables). 
A lot of future work seems called for.

\bibliography{main}
\bibliographystyle{splncs03}

\end{document}